\documentclass[aoas,nameyear,seceqn,dvips]{arximspdf}
\usepackage{graphics,textcomp}
\doi{10.1214/09-AOAS253}
\volume{3}
\issue{4}
\pubyear{2009}
\firstpage{1467}
\lastpage{1492}

\makeatletter
\newcommand{\mr}[1]{\mathrm{#1}}
\newcommand{\mf}[1]{\mathbf{#1}}

\def\cac{$\mr{Ca}^{2+}$}
\makeatother

\begin{document}
\begin{frontmatter}

\title{Use of multiple singular value decompositions to analyze complex
intracellular calcium ion signals}
\runtitle{SVD of calcium ion signals}

\begin{aug}
\author[a]{\fnms{Josue G.} \snm{Martinez}\thanksref{t1}\ead[label=e1]{jgmartinez@stat.tamu.edu}},
\author[a]{\fnms{Jianhua Z.} \snm{Huang}\thanksref{t2,t3}\ead[label=e2]{jianhua@stat.tamu.edu}},
\author[b]{\fnms{Robert C.} \snm{Burghardt}\thanksref{t4}\ead[label=e3]{rburghardt@cvm.tamu.edu}},
\author[b]{\fnms{Rola} \snm{Barhoumi}\thanksref{t4}\ead[label=e4]{rmouneimne@cvm.tamu.edu}}
\and
\author[a]{\fnms{Raymond J.} \snm{Carroll}\thanksref{t2}\ead[label=e5]{carroll@stat.tamu.edu}\corref{}}
\runauthor{J. G. Martinez et al.}
\affiliation{Texas A\&M University}
\address[a]{
J. G. Martinez\\
J. Z. Huang\\
R. J. Carroll\\
Department of Statistics\\
Texas A\&M University\\
3143 TAMU\\
College Station, Texas 77843-3143\\
USA\\
E-mails: \printead*{e1}\\
\phantom{E-mails: }\printead*{e2}\\
\phantom{E-mails: }\printead*{e5}}
\address[b]{R. C. Burghardt\\
R. Barhoumi\\
Department of Veterinary\\
\quad Integrative Biosciences\\
Texas A\&M University\\
4458 TAMU\\
College Station, Texas 77843-4458\\
USA \\
E-mails: \printead*{e3}\\
\phantom{E-mails: }\printead*{e4}}

\thankstext{t1}{Supported by a postdoctoral training grant from the
National Cancer Institute (CA90301).}
\thankstext{t2}{Supported by a grant from the National Cancer Institute
(CA57030) and by Award Number KUS-CI-016-04, made by King Abdullah
University of Science and Technology (KAUST).}
\thankstext{t3}{Supported by a grant from the National Science
Foundation (DMS-06-06580).}
\thankstext{t4}{Calcium imaging performed in the College of Veterinary
Medicine \& Biomedical Sciences Image Analysis Laboratory, was
supported by NIH-NIEHS Grants P30-ES09106, P42-ES04917 and T32 ES07273.}
\end{aug}

\received{\smonth{12} \syear{2008}}
\revised{\smonth{4} \syear{2009}}

%
\begin{abstract}
We compare calcium ion signaling (\cac) between two exposures; the data
are present as movies, or, more prosaically, time series of images.
This paper describes novel uses of singular value decompositions (SVD)
and weighted versions of them (WSVD) to extract the signals from such
movies, in a way that is semi-automatic and tuned closely to the actual
data and their many complexities. These complexities include the
following. First, the images themselves are of no interest: all
interest focuses on the behavior of individual cells across time, and
thus, the cells need to be segmented in an automated manner. Second,
the cells themselves have 100$+$ pixels, so that they form 100$+$ curves
measured over time, so that data compression is required to extract the
features of these curves. Third, some of the pixels in some of the
cells are subject to image saturation due to bit depth limits, and this
saturation needs to be accounted for if one is to normalize the images
in a reasonably unbiased manner. Finally, the $\mr{Ca}^{2+}$ signals have
oscillations or waves that vary with time and these signals need to be
extracted. Thus, our aim is to show how to use multiple weighted and
standard singular value decompositions to detect, extract and clarify
the $\mr{Ca}^{2+}$ signals. Our signal extraction methods then lead to simple
although finely focused statistical methods to compare $\mr{Ca}^{2+}$ signals
across experimental conditions.
\end{abstract}

%
%

\end{frontmatter}
%
\section{Introduction}\label{sec1}

Scientifically, this paper is about the study of the effects
of~2,3,7,8-Tetrachlorodibenzo-p-dioxin (TCDD) on calcium ion signaling
(\cac) in myometrial cells. The importance of $\mr{Ca}^{2+}$
signaling in cell
function, for example, metabolism, contraction, cell death,
communication, cell proliferation, has been studied in numerous types
of cells; see Putney (\citeyear{Putneyb8}). TCDD itself is a toxicant
by-product of
incomplete combustion of fossil fuels, woods and wastes and is known to
adversely effect reproduction, development and the immune system as
well as being a probable carcinogen.

The essential feature of these data is that they present themselves as
movies of~512 images, or time series of images after oxytocin exposure.
To best appreciate the complexity of the data, and thus this paper,
readers should first look at two of the movies, in the \hyperref
[suppA]{Supplementary Materials}, 
one without and one with TCDD exposure.

The experiment leading to these images is described in detail in
Section \ref{data}. However, the movies show that the data are complex,
and analysis of them is not simple. This paper describes novel uses of
singular value decompositions (SVD) and weighted versions of them
(WSVD) to extract the signals from such movies, in a way that is
semi-automatic and tuned closely to the actual data and their many
complexities. Here we describe a few of these complexities:
\begin{enumerate}[]
\item[\textit{Basic background}.] The data consist of 512 images. Myometrial
cells can
be seen in these images, which start out in their native state and are
then exposed to an oxytocin stimulus, at which point $\mr{Ca}^{2+}$ expression
becomes pronounced. The cells themselves are fixed to a substrate and
do not move over time.
\item[I. \textit{Cell segmentation}.] The images themselves are of no intrinsic
interest:
what matters is how the individual cells express \cac. This means that
segmenting the image to obtain the cells is a crucial first step. To
see what has been done in the past, consider Figure \ref
{fig:calspike-lowT}, which gives a sequence of images in the first 2
minutes of the experiment. Because it is difficult to distinguish cell
boundaries before oxytocin is delivered, it is common to use a static
approach. Specifically, the brightest image is used to isolate the
cells, with cell boundaries drawn by hand. This technique, although
practical, is not semi-automatic and uses only a small fraction of the
information available because it ignores the 511 other images that
could have pertinent information about the cell boundary. This could
potentially lead to under or overestimation of the cell boundaries.
Instead, we will describe a method that allows use of all 512 images in
order to determine cell location. Our approach utilizes the brightest
image to get a rough idea of the cell location and then obtains a
summary of the resulting pixel-wise matrix of all 512 images to refine
the cell boundaries.

%
\begin{figure}

\includegraphics{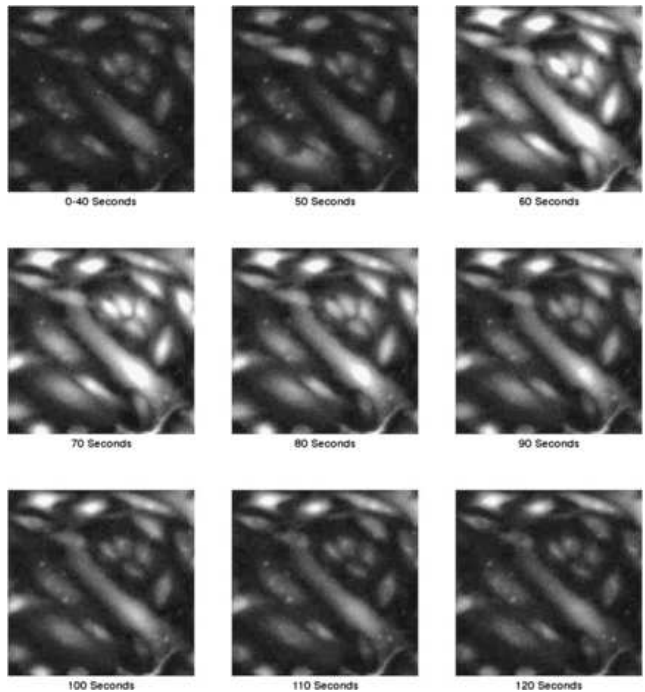}

\caption{Oxytocin-induced calcium response in myometrial cells during the
first 2 minutes of the experiment. Cells were cultured in a low
level of estrogen/progesterone and were treated with 10~nM TCDD
for 24 hours. Cells were then loaded with the Fluo-4, washed and then
stimulated with 20~nM oxytocin following identification of basal
calcium levels in cells. The movie of this cell line as well as the nontreated
one cultured in low hormone level are available as part of the \protect
\hyperref[suppA]{Supplementary Materials}.}
\label{fig:calspike-lowT}
\end{figure}
\item[II. $\mr{Ca}^{2+}$ \textit{signal extraction}.] Each segmented cell will
contain 100$+$ pixels, and
each of these pixels is its own movie or curve. Immediately, one is
faced with the problem of summarizing these curves. The usual choice of
a summary statistic in $\mr{Ca}^{2+}$ signal publications is the normalized
average signal across time, where ``normalized'' means that the whole
signal is divided by the initial signal values recorded before the
stimulus is delivered to the cells; see Barhoumi et al. (\citeyear
{Barhoumib2}) and
Burghardt et al. (\citeyear{Burghardtb4}). Hence, signal amplitude is
measured in units
of ``fold change,'' compared to the $\mr{Ca}^{2+}$ signal before
stimulus. While
convenient, this method ignores the potential for additional
information in the wealth of pixel information, information we aim to
extract, and it is in addition not necessarily the best way to
normalize the data.

\item[III. $\mr{Ca}^{2+}$ \textit{signal clarification}.] Having extracted the
basic signal, we face a further
obstacle. An unusual feature of these data is that some of the pixels
in most of the cells reach image saturation. This type of image
censoring has the potential to distort downstream statistical analysis,
is generally ignored in the literature, and needs to be accounted for.
That is, we wish to clarify the original signal to account for image saturation.

\item[IV. $\mr{Ca}^{2+}$ \textit{treatment comparisons}.]
Having segmented the cells, and extracted and
clarified the cell $\mr{Ca}^{2+}$ signal, we are then in a position to
understand
some of the effects of TCDD exposure.

\item[$\mr{Ca}^{2+}$ \textit{information extraction is the key}.] The main
point of this paper is to extract the
information in the movies, in a semi-automatic way that reduces the
potential for bias.

\item[$\mr{Ca}^{2+}$ \textit{singular value decompositions}.] In this paper we
will show how to use the singular
value decomposition (SVD) and a novel weighted singular value
decomposition (WSVD) to perform the four crucial steps I--IV. Each step
requires a different use of the SVD or WSVD. We demonstrate that novel
uses of SVD/WSVD help us understand the effect of TCDD exposure.
\end{enumerate}

Our paper is organized in the following manner. In Section \ref{data}
we describe
the experiment and the data. We proceed to restate the singular value
decomposition (SVD) in Section \ref{sec:data-anl} and demonstrate how
to use it to obtain
the EigenPixel and EigenSignal vectors. In Section \ref{sec:data-acq}
we outline the use
of the SVD to detect the $\mr{Ca}^{2+}$ signal from images, that is,
to segment
the cells. In Section \ref{sec5} we use yet another SVD
to isolate the $\mr{Ca}^{2+}$ signal from the resulting pixel-wise matrices.
We implement a weighted SVD, WSVD, with a clever choice of weights in
Section \ref{sec:wsvd}, and use it to remove
the saturation effect on the $\mr{Ca}^{2+}$ signal. Finally, in
Section~\ref{sec:comp} we compare
the control and treated cells by applying the SVD once more to obtain
one point summary values for each cell, that is, EigenCells, which
enable us
to distinguish between control and treated groups. We offer some concluding
remarks in Section \ref{sec:conclusion}.

\section{Experiment}\label{data}
\subsection{Introduction}
The essential statistical details of this experiment are that there are
myometrial cells fixed to different substrates, one of which is exposed
to TCDD and the other of which is not. Shortly after image capturing
commences, the cells are exposed to oxytocin, thus stimulating the $\mr
{Ca}^{2+}$
signal. The main goal is to compare the TCDD exposure to the control.
What follows are some of the details of the experiment.

\subsection{Treatments}
Myometrial cells, which comprise the contractile middle layer of the
uterine wall, were cultured in three levels of an estrogen/progesterone
hormone combination: basal, low and high. The ``basal'' level is the one
in which the cells were cultured, the ``low'' level of hormone is
slightly higher than that found in women before pregnancy and the
``high'' level is the level of a pregnant woman at full term. Our work
presents data from two different treatments (control or TCDD) with 3
different levels of hormones in the culture medium (basal, low and high).

The treated cells received a 100 nM solution of TCDD 24 hours before
the experiment. Cells are cultured on coverglass chambered slides. All
cells were then washed and loaded with 3 \textmu M Fluo-4 for 1 hour at
$\mr{37^{\circ}C}$: fluorescent probe Fluo-4 is one of many dyes used to
detect changes in $\mr{Ca}^{2+}$ within cells. Fluo-4 is typically
excited by
visible light of about 488 nM, and emits about 100 fold greater
fluorescence at about 520 nM upon binding free $\mr{Ca}^{2+}$. Following loading,
cells were washed and placed on the stage of the confocal microscope.
Cells were then scanned five times to establish the basal level of $\mr
{Ca}^{2+}$
prior to addition of 20 nM oxytocin, the hormone used in this study to
stimulate $\mr{Ca}^{2+}$ signal in these cells. Scanning continues at
10 second
intervals for approximately 85 minutes, leading to 512 images (100
$\times$ 100 pixels) containing 20--50 cells per treatment.

\subsection{Imaging}

The data captured in these experiments are digital images of~$\mr{Ca}^{2+}$
fluorescence of individual cells. The bit depth of images used in this
study is of 8 bits, which translates to $2^8$ or 256 possible grayscale
values in the image. Unfortunately, it often happens that the maximum
concentrations detected in these images are limited by the bit depth.
This may sometimes result in saturation and lead to underestimation of
changes in $\mr{Ca}^{2+}$ signals, especially when multiple treatments are
performed and accurate evaluation of these differences is required.

Figure~\ref{fig:calspike-lowT} shows a response to the oxytocin
stimulus, in cells treated with TCDD and cultured in a low
estrogen/progesterone hormone level. The maximal reaction due to the
oxytocin challenge appears at 60 seconds and then the cells return to
their steady state. Notice that not all cells go back to their steady
state at the same rate. In fact, there is residual fluorescence in some
cells at the top of each of the images in Figure~\ref
{fig:calspike-lowT}, long after the initial peak of fluorescence at 60 seconds.

\subsection{Overview of what is to come}
In order to study the intracellular $\mr{Ca}^{2+}$ signal, we make use
of the
singular value decomposition in four ways. First we isolate or (a)
detect the cell itself. To do this we perform a singular value
decomposition on a matrix made up of pixels from a rough segmentation
of each cell. The spatial plot of the first EigenPixel resulting from
this SVD is used to determine which pixels are important when
harnessing the signal. The next step
is to (b) extract the~$\mr{Ca}^{2+}$ signal. In this step we apply the
SVD on the
resulting pixel-wise matrix from the previous step and obtain the first
EigenSignal, which contains most of the~$\mr{Ca}^{2+}$ signal
information of the
cell of interest. The third step is to (c) clarify that signal. In this
step we adapt the usual SVD and introduce a weighted SVD which takes
care of two problems: (1) it imputes
values where pixel saturation occurs and (2) it weights the influence of
each pixel based on variance. Finally, the last step in our study of
intracellular $\mr{Ca}^{2+}$ signal is to (d) compare the effect of
the carcinogen
TCDD across experimental conditions to see how it affects the $\mr{Ca}^{2+}$
response. To accomplish this, we use the SVD again to obtain one point
summary values for each cell.
\section{SVD after rough segmentation}
\label{sec:data-anl}

\subsection{Outline}
This section describes the well known SVD and outlines part of how we
will use it, after large rectangular regions containing each cell have
been obtained (rough segmentations). We particularly need to describe
some terminology for future use.

\subsection{EigenPixels and EigenSignals}

The singular value decomposition (SVD) is a widely used matrix
factorization technique. For example, the SVD was used to analyze
microarray expression data, where the rows of the matrix in
question comprise the genes and the columns represent the
expression arrays [Alter et al.~(\citeyear{Alterb0})]. This use of the
SVD introduced
the idea
of transforming the gene, array space to an ``eigengene,'' ``eigenarray''
space that is reduced and diagonalized. We will
draw inspiration from this approach and show that we can use
``eigen $\mr{Ca}^{2+}$ signal'' or EigenSignal vectors to summarize
the $\mr{Ca}^{2+}$
response for each cell in the experiment and we will later
describe how we acquired matrix representations for each cell.

We first describe how to obtain ``eigen pixel'' and
``eigen $\mr{Ca}^{2+}$ signal'' vectors, using the SVD. To accomplish this,
we will present the singular value decomposition in the context
of our data, assuming that a rough segmentation of the cells has been
performed. For all treatments considered in this work, we
represent each cell as a matrix of $\mr{Ca}^{2+}$ intensity, in grayscale
values, that has a number of pixels which comprise the cell,
for all 85 minutes of the experiment. Each matrix has $n$ rows
and $m$ columns, where $n$ is the number of pixels that represent
the cell and $m$ is the number of time points in the experiment.
All cells were observed the same number of times so $m=512$.
Let $X_{k}$ represent the $n \times m$ calcium signal matrix for
the $k$th-cell. The singular value decomposition of $X_{k}$ is
%
%
\begin{equation}\label{eq:svd}
X_{k}=U_{k}S_{k}V_{k}^{\mathrm{T}}.
\end{equation}
Here $V_{k}$ is an $m \times n$ matrix whose column vectors,
$\mathbf{v}_{kj}\in\mathbb{R}^{m}$, form an orthonormal basis
for the $\mr{Ca}^{2+}$ signal, and are called EigenSignal vectors.
In (\ref{eq:svd}) $U_{k}$ is an $n \times n$ matrix whose column
vectors, $\mathrm{u}_{kj}\in\mathbb{R}^{n}$, form an orthonormal
basis for the pixels of the cell, called EigenPixel vectors.
In addition, $S_{k}$ is an $n \times n$ square matrix of singular
values arranged from largest to smallest
$s_{k1} \ge s_{k2} \ge\cdots\ge s_{kn}$.

We can generate a rank-$L$ matrix that approximates $X_{k}$ by
using the first $L$ $u_{kj}$ and $v_{kj}$ vectors, that is,
%
%
\begin{equation}\label{eq:lowrank}
X_{k}^{L}=\sum_{j=1}^{L}u_{kj}s_{kj}v_{kj}^{\mathrm{T}}.
\end{equation}
In equation (\ref{eq:lowrank}) $X_{k}^{L}$ is the best
rank-$L$ matrix that approximates $X_{k}$, in the sense that it
minimizes the sum of squares difference between $X_{k}^{L}$
and $X_{k}$ among all rank-$L$ matrices [Trefethen and Bau (\citeyear
{Trefethenb12})].
Low rank approximations are useful because less data are needed
to represent the original matrix; these techniques are often
used in image compression. We will use the smallest
number of EigenPixel and EigenSignal vectors that summarize
both pixel and $\mr{Ca}^{2+}$ signal information.


\section{$\mr{Ca}^{2+}$ cell segmentation}
\label{sec:data-acq}

\subsection{Peak image}

The cells used in this study are cultured as monolayer on coverglass
chambered slides. This allows easy imaging of the cells over time
without any movement: the cells in this study are fixed in a substrate.
This fact is essential to the work that follows.

As may be apparent from the sequence of images shown in Figure \ref
{fig:calspike-lowT}, it is difficult to distinguish cell boundaries
before oxytocin is delivered. For this reason, in order to determine
the location of the cells, as well as their boundaries, it is common to
use the brightest image to isolate the cells. This technique, although
practical, only uses a small fraction of the information available
because it ignores the 511 other images that could have pertinent
information about the cell boundary. Instead, we propose that a summary
of these 512 images should be used to determine cell location. Our
approach makes use of the brightest image, or ``peak'' image, to get a
rough idea of the cell location and then uses a summary of the
resulting pixel-wise matrix of all 512 images to refine the cell
boundary that will be used for the rest of the analysis. We use the
image where we see the most distinction between cell boundaries as the
``peak'' image.

%
\begin{figure}[b]

\includegraphics{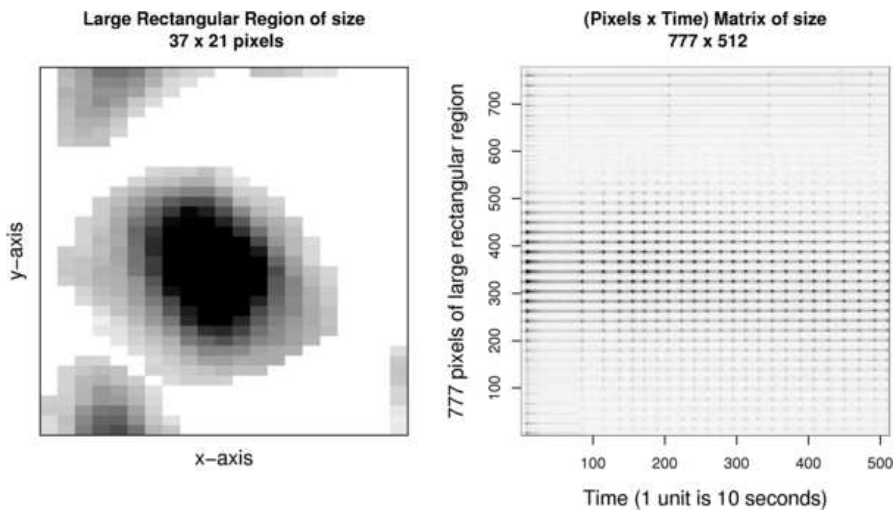}

\caption{The initial rough rectangular segmentation
of cell 2 from the treated group of low hormone level and the
corresponding 777 $\times$ 512 pixel-wise matrix for this rectangular region.}
\label{fig:rough}
\end{figure}

\subsection{$\mr{Ca}^{2+}$ signal detection via first eigenpixel}

\begin{figure}

\includegraphics{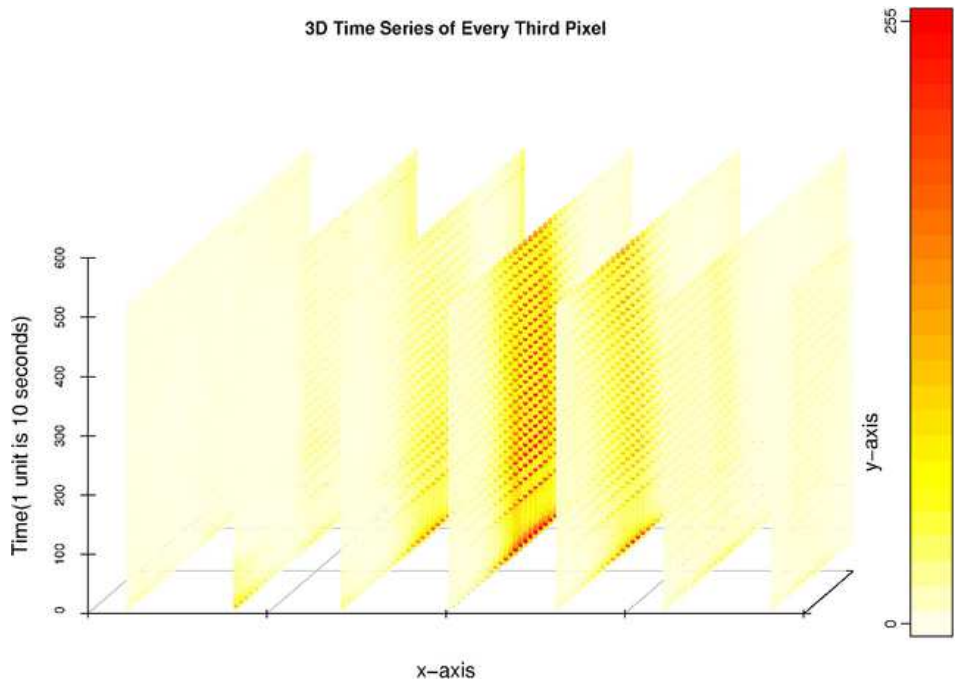}

\caption{3D plot of the $\mr{Ca}^{2+}$ intensity in the rough
segmented region of cell 2 from the treated group of low hormone level,
corresponding to the left panel of Figure~\protect\ref{fig:rough}. The
$x$--$y$
coordinates correspond to space and the vertical coordinate to time.
Every fourth pixel is shown.}
\label{fig:rough3d}
\end{figure}

Once the ``peak'' image from each cell line is identified, we draw very
large rectangular regions each containing a cell. Each rectangular
region assures that the boundaries of the cell of interest are
contained within it, although there may be parts of other cells that
fall in this rectangular region. Figure~\ref{fig:rough} shows the
rectangular region chosen from the ``peak'' image to represent the
rough segmentation of cell 2, from the treated group in the low hormone
level. Figure~\ref{fig:rough} also displays the resulting 777 $\times$
512 pixel-wise matrix derived by taking the 777 pixels that represent
the rectangular region from each distinct image at every one of the 512
time points. The right panel of Figure~\ref{fig:rough} does not respect
the spatial location of the pixels. A better view of how the
3-dimensional time series of $\mr{Ca}^{2+}$ intensity evolves is shown
in Figure~\ref
{fig:rough3d}. This perspective plot of every third pixel in the rough
segmentation shows the spatial location of pixels over time. Notice
that the oscillations in the signal concentrate in the center of the
$x$--$y$ plane and evolve over time in the $z$-axis.

\begin{figure}

\includegraphics{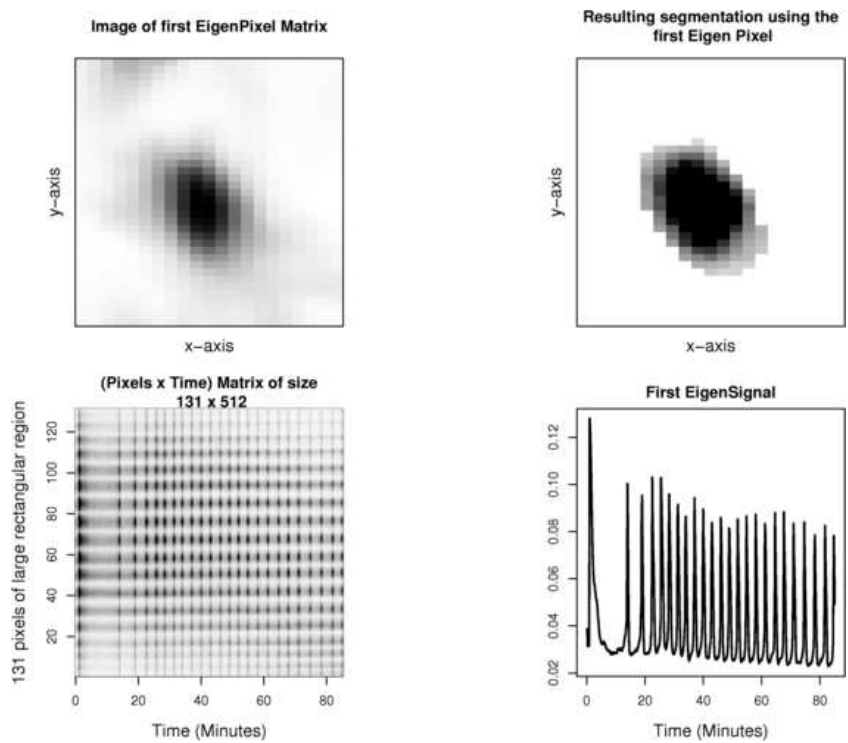}

\caption{Top row: Image of the first
EigenPixel vector obtained from the SVD of the rough \mbox{777 $\times
$ 512}
pixel-wise matrix and the resulting segmentation of cell 2 after using
the first EigenPixel to perform the segmentation.
Bottom row: The corresponding 131 $\times$ 512 pixel-wise
matrix for this new segmentation and the corresponding first
EigenSignal over the 85 minute experiment.}
\label{fig:rough2}
\end{figure}

If $X_{2}$ represents the 777 $\times$ 512 pixel-wise matrix of pixels
$\times$ time for cell 2, shown in the right panel of Figure \ref
{fig:rough}, then we obtain a summary of the pixel information by
taking the SVD of $X_{2}$ and obtaining the first EigenPixel. As
explained in Section \ref{subsec:5.2} below, only the first singular
value explains the majority of the variance in these data, hence, the
first EigenPixel summarizes all the pixel information to one vector of
size 777. We take this vector and plot it spatially on the
corresponding pixel location. What we get is a 2-dimensional image
where the pixel intensity reflects the importance of the pixel in
representing the $\mr{Ca}^{2+}$ signal of this cell (top left panel of
Figure \ref
{fig:rough2}). This image is a better candidate for use in
identification of the $\mr{Ca}^{2+}$ signal than the ``peak'' image
because it
summarizes the importance of each pixel across the 512 images in the
experiment. This is our first use of the SVD and the way in which we
will detect the $\mr{Ca}^{2+}$ signal for all cells in this experiment.

%

%

%
\begin{figure}

\includegraphics{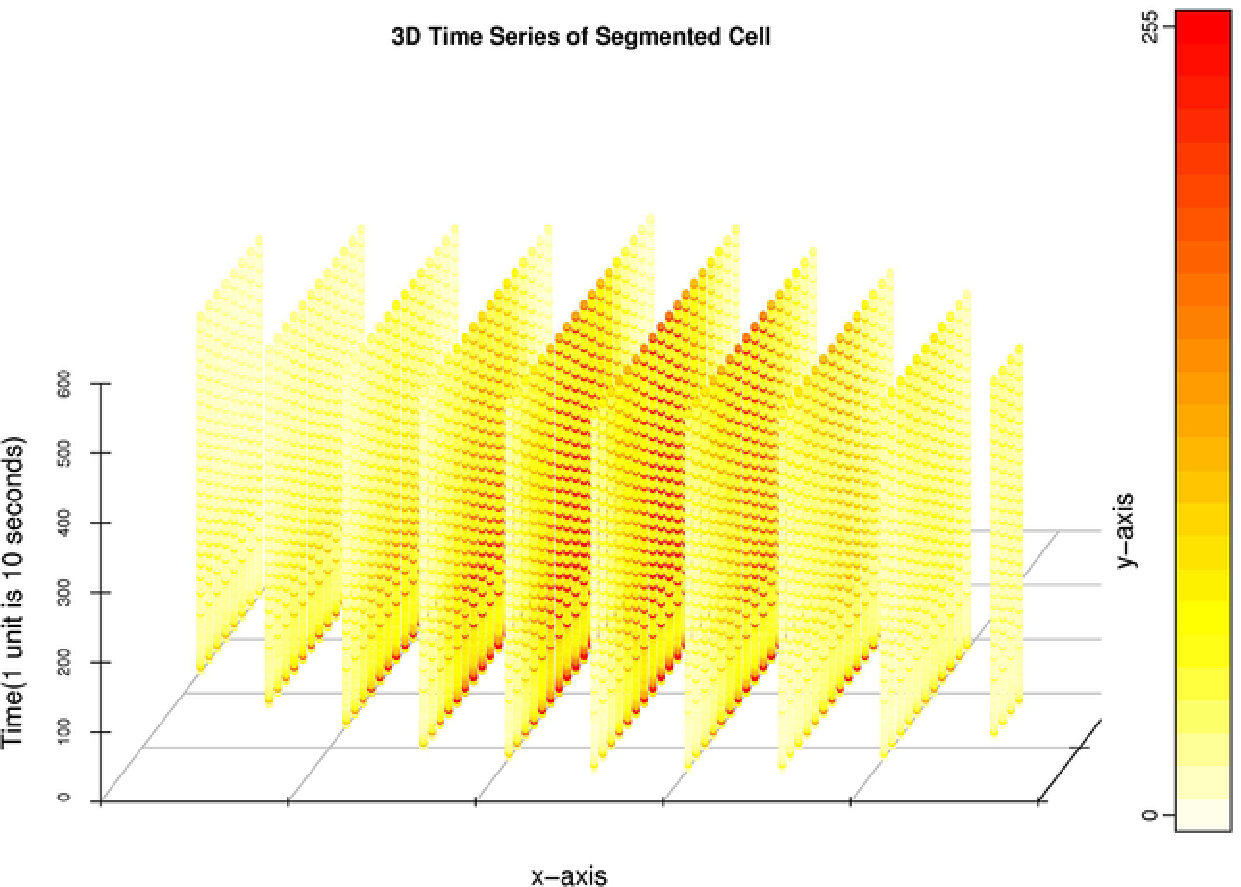}

\caption{3D plot of the $\mr{Ca}^{2+}$ intensity for the final
segmentation of cell 2 from the treated group of low hormone level,
corresponding to the left panel of Figure~\protect\ref{fig:rough2}.
The x--y
coordinates correspond to space and the vertical coordinate to time.
Every fourth pixel is shown.}
\label{fig:rough23d}
\end{figure}

%
\subsection{$\mr{Ca}^{2+}$ final segmentation}

Once we obtain this first EigenPixel image from~$X_{2}$, we
use the EBImage package from Bioconductor to segment and index the cell
[R Development Core Team (\citeyear{RDevb9})].
We first blurred the image to smooth out any noisy pixels. We
then used thresholding to pick out the region of high pixel
values which usually contains the cell, and finally used a
watershedding algorithm to close the cell boundaries and separate
other cell chunks that are close together. The result is the final
segmentation of the cell shown in the top right panel of Figure \ref
{fig:rough2}.
Notice that all we have done is pick the region with highest EigenPixel
intensity
which in turn should give us the spatial location of the pixels that
contain most of the $\mr{Ca}^{2+}$ signal information. We then collect
each of the
131 pixels
in this final segmentation from each of the 512 images and get a matrix
representation of the cell;
see the bottom left panel of Figure \ref{fig:rough2}. As before, this
matrix does not
respect the spatial location of the pixels, hence, we provide a
3-dimensional plot where
each of the pixels in the final segmentation is displayed over time;
see Figure \ref{fig:rough23d}.
It is easier to appreciate the spatial pattern of the $\mr{Ca}^{2+}$
signal and
where it concentrates on the $x$--$y$ plane
at any time.

We used this segmentation process to generate contours of each cell,
and used these contours to pick
out the cell position from every image at every one of the~512 time
points. This process yielded 20--50 cells from each treatment.

The oscillatory behavior observed in Figures \ref{fig:rough2} and \ref
{fig:rough23d} and
throughout the text are present because calcium ions ($\mr{Ca}^{2+}$)
are responsible
for many important physiological functions. In smooth muscle cells that
surround hollow organs of the
body, transient increases in intracellular $\mr{Ca}^{2+}$ can be
stimulated by a
number of hormones to activate smooth muscle contraction. Because
sustained elevation of~$\mr{Ca}^{2+}$ is toxic to cells, $\mr
{Ca}^{2+}$ signals in many cell
types frequently occur as repetitive increases in $\mr{Ca}^{2+}$,
referred to as
$\mr{Ca}^{2+}$ oscillations. The periodic $\mr{Ca}^{2+}$ spikes which
increase with
increasing hormone concentration are thought to constitute a~frequency
encoded signal with a high signal-to-noise ratio which limits prolonged
exposure of cells to high intracellular $\mr{Ca}^{2+}$; see Sneyd,
Keizer and Sanderson~(\citeyear{Sneydb11}).
Interestingly, the frequency of $\mr{Ca}^{2+}$ oscillations in smooth
muscle cells
is relatively low (e.g., 2--10 MHz) [see Burghardt et al.~(\citeyear
{Burghardtb4})], whereas
in liver cells which use $\mr{Ca}^{2+}$ oscillations to stimulate ATP
production
in mitochondria and the breakdown of glycogen to glucose, the frequency
of $\mr{Ca}^{2+}$ oscillations is much greater (e.g., range from 5 to
100 MHz);
see Barhoumi et al.~(\citeyear{Barhoumib2}). The spatial and temporal
organization and the
control of these intracellular $\mr{Ca}^{2+}$ signals is of
considerable interest
to cellular biologists.

\section{Signal extraction}\label{sec5}

\subsection{Overview}
The top right panel of Figure \ref{fig:rough2}
shows the region that represents cell 2 and the bottom left panel
shows the resulting 131 $\times$ 512 pixel-wise matrix for the final
segmentation, which we will label as $X^{\prime}_{2}$. We then use
the singular value decomposition once again and obtain the first
EigenSignal from the $X^{\prime}_{2}$ matrix shown in the bottom right
panel of Figure \ref{fig:rough2}; this is our $\mr{Ca}^{2+}$ signal extraction
step. The $\mr{Ca}^{2+}$ signal produced from this step is a candidate signal
that represents a summary of the $\mr{Ca}^{2+}$ intensity for the cell
in question.

\subsection{First EigenPixel and EigenSignal}\label{subsec:5.2}

\begin{figure}

\includegraphics{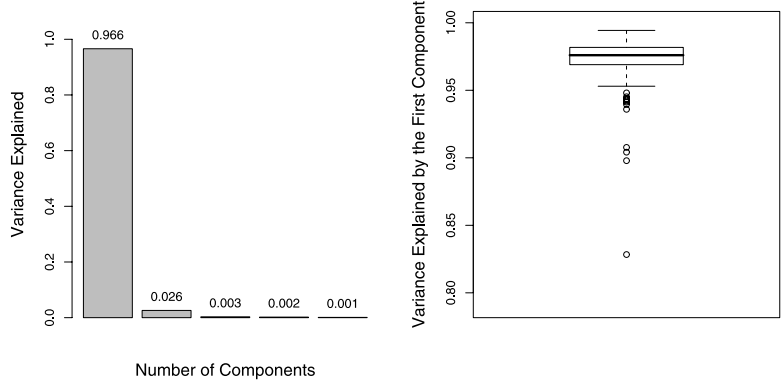}

\caption{Left: Variance explained by the
first 5 components in the SVD
of the pixel-wise matrix that represents one cell \textup{(11)} from
the control group in the high hormone treatment. Right: Variance
explained by the first component in the SVD in each of the 187
cells examined in all six treatments used in the study.}
\label{fig:impcomp}
\end{figure}

When we take the SVD of each matrix for each cell, in all cell lines,
we find that the first EigenSignal vector is enough to give a good
representation of the $\mr{Ca}^{2+}$ signal in these data, because the first
singular value basically dominates the signal in the data. In fact, if
we take the ratio of first to second singular values for each cell and
take the mean, we find that on average the first singular value is
between 8 to 10 times larger than the second, and many times larger
than the $3$rd and $4$th singular values. The left panel of Figure \ref
{fig:impcomp} shows the variance explained by the first five singular
values of the SVD of cell 11 in the control group of the high hormone
level cell line, where the first component explains 97\% of the
variance. On average, the variance explained by the first component in
each of the 187 cells considered across all cell lines in this
experiment is 97\%. The right panel of Figure \ref{fig:impcomp} shows
the distribution of the variance explained by the first component for
each of the 187 cells. The minimum variance explained by the first
singular value among the 187 cells is 83\%, hence the first EigenSignal
and EigenPixel vectors that correspond to this first singular value
summarize almost all the $\mr{Ca}^{2+}$ signal and pixel information
in each of
these matrices. For this reason we will assume that only the first
EigenSignal and first EigenPixel are needed to summarize the $\mr
{Ca}^{2+}$ signal
and pixel information in the data.
%
%

\section{$\mr{Ca}^{2+}$ signal clarification and cell saturation}
\label{sec:wsvd}

\subsection{The problem of saturation}
The saturation phenomenon is based upon the fluorescence
detection system. The fluorescence detection system utilized in experiments
presented in this report is a photomultiplier detector tube (PMT).
This detector does not count individual photons; rather it requires a
certain minimum number of
photons to activate an electrode which will emit a small number of
electrons which are subsequently amplified in a stepwise fashion. The
readout of the PMT is on a 256 grey scale level. Occasionally the
amplified signal can reach saturation if the fluorescence output of the
calcium signal being detected is very high. Normally, the settings of
the PMT are adjusted so as not to reach saturation, however, detection
of the low end of the fluorescence signal is very important.

\begin{figure}

\includegraphics{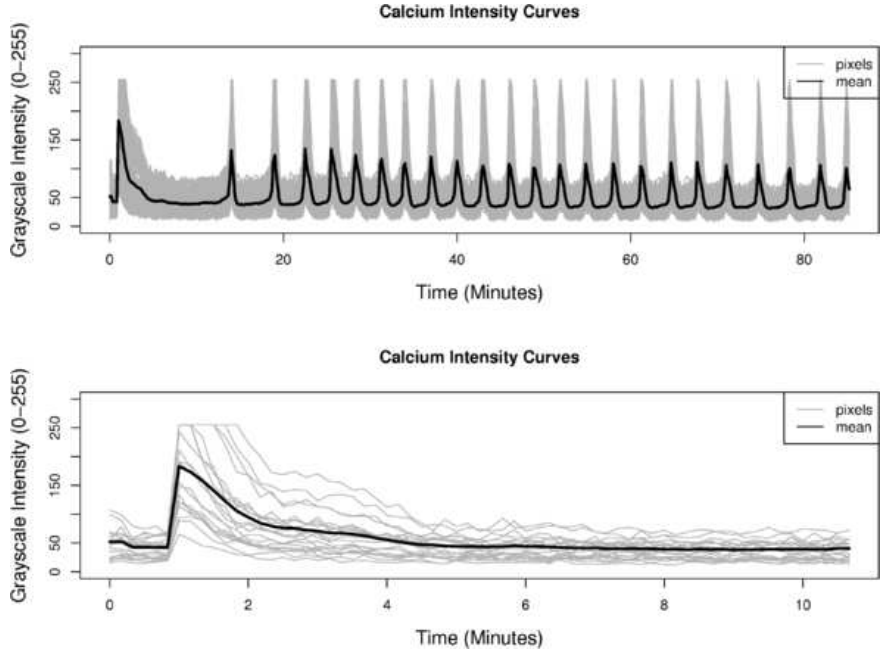}

\caption{Top: The Calcium intensity curves
over the 85 minute experiment of the 131 pixels in the $131 \times512$
pixel-wise matrix $X^{\prime}_{2}$. Bottom: 20 randomly selected
pixels from $X^{\prime}_{2}$.}
\label{fig:rough3}
\end{figure}

Notice that the grayscale values of some of the pixels that
represent cell 2, shown in Figure \ref{fig:rough2}, reach a ceiling of
255; see Figure \ref{fig:rough3}.
This is especially noticeable after the cell received the oxytocin
stimulus around 1 minute into the experiment. Because the individual pixel
values reflect the $\mr{Ca}^{2+}$ level in the cell, $\mr{Ca}^{2+}$
summary measures will
undoubtedly be affected if the pixels reach the ceiling of 255. Also notice
the variability in individual pixel values. The bottom panel of Figure~\ref{fig:rough3} shows
the intensity of 20 pixels over time and it is clear that some may
reach maximum intensity values
that are larger than 255 and some at much lower values. We do not model
the behavior of individual pixels
in this work but it can certainly be considered in the future.

Two questions are immediate. First, how is the EigenSignal affected
when pixel values reach the saturation level of 255? Second, how
should one process the $\mr{Ca}^{2+}$ signal once pixel saturation has been
detected? We implement the algorithm introduced by Gabriel and Zamir
(\citeyear{Gabrielb5}) and
let the saturated pixels be missing data to address this issue.
To our knowledge, there are no methods in the literature
available to deal with the clarification of $\mr{Ca}^{2+}$ signal curves
and our attempt is the first of its kind.
%

\subsection{The weighted SVD}

\begin{figure}

\includegraphics{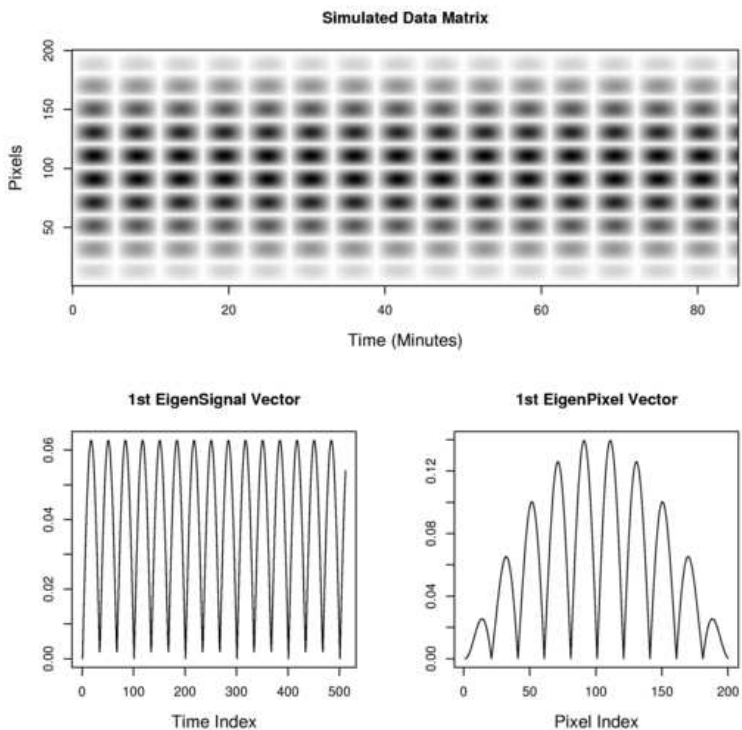}

\caption{Top row: Simulated data matrix
(before adding noise)
to be tested. Bottom row:
First EigenSignal and EigenPixel curves obtained from the data matrix
shown above.}
\label{fig:sdp}
\end{figure}

\begin{figure}

\includegraphics{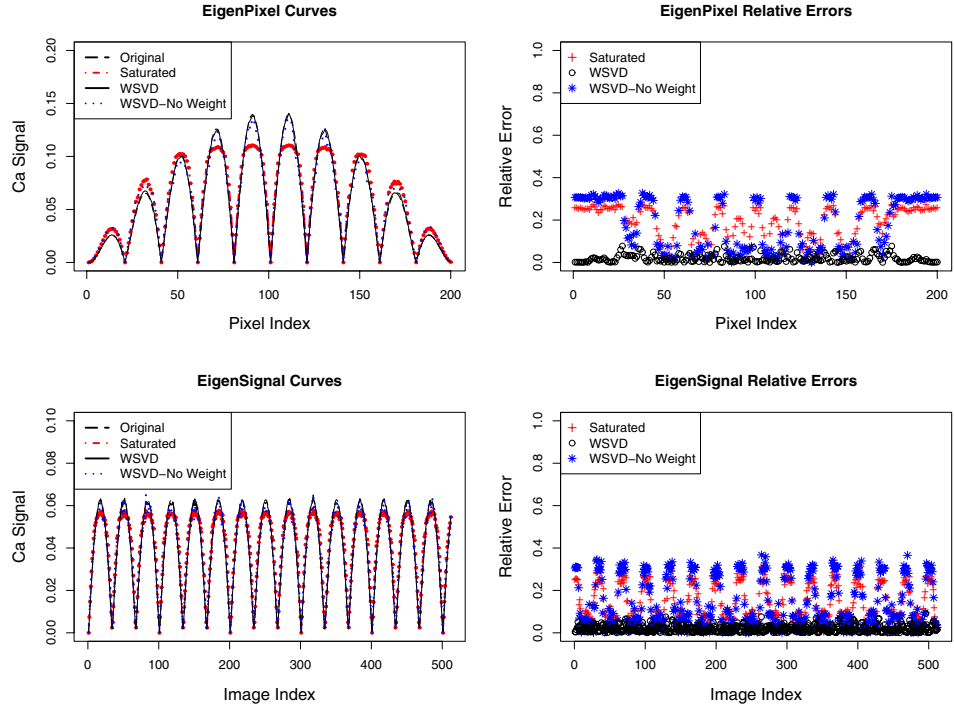}

\caption{Top row:
EigenPixel vectors of the original, saturated, weighted SVD (WSVD)
and WSVD with no weights (WSVD-No Weight)
and the relative error of the saturated and final WSVD and WSVD-No
Weight EigenPixels.
Bottom row: EigenSignal vectors of the original, saturated, weighted
SVD (WSVD)
and WSVD with no weights (WSVD-No Weight) and the corresponding error
curves of the
saturated and final WSVD and WSVD-No Weight EigenSignals.
}
\label{fig:wsvd}
\end{figure}

Although the first EigenSignal is a reasonable measure to use when
summarizing the $\mr{Ca}^{2+}$ signals of the pixel-wise matrices,
there are
drawbacks if used without adjustment. If there are too many pixels that
reach the saturation point, the signal can be under or over-estimated
at different time points in the experiment and a misinterpretation of
the signal amplitude can occur. It is intuitive to understand how the
signal can be under-estimated due to saturation, but over-estimation of
the summary signal is certainly an unexpected phenomenon which we will
explain. Because pixels reach a saturation, the signal summary is
undoubtedly affected by this lack of information on maximal values
attained by such pixels. In the course of the time series where many
pixels reach saturation, the signal is under-estimated around the
peaks, but the effect of this under-estimation results in an
over-estimation during a time where no pixels reached saturation.
Figures \ref{fig:sdp} and \ref{fig:wsvd} show this phenomenon.
The over-estimation is due to the normalization requirement of the
singular vectors and the right skewness of the cross-sectional
intensity distribution.

To correct these over- and under-estimation effects, we must remove the
effect of the saturated pixels and recalculate the $\mr{Ca}^{2+}$
signal without
their influence. One approach is to simply remove every row of the
pixel-wise data matrix which contains a saturated pixel and recompute
the $\mr{Ca}^{2+}$ signal using the resulting matrix; however, this
could lead to
the removal of a significant number of rows from the data matrix.
Instead, we propose to implement the weighted SVD, WSVD, using the low
rank matrix approximation of Gabriel and Zamir (\citeyear{Gabrielb5})
where we
introduce the use of indicators in the weights, as in Beckers and Rixen
(\citeyear{Beckersb3}), to treat the saturated pixels as missing data
and use a clever
choice of weights that allows for accurate recovery of the original signal.

It is important to note that our ``missing data'' is not really
missing, we know that the saturated pixels must at least attain a value
of 255. Hence, if we observe values that are below this threshold in
our imputation, we would certainly know that we've made an error. We
implement a check in our algorithm that gives us a~flag if a value that
is initially saturated falls below its saturation point.

Imputation of missing values in the SVD is not a new subject. As noted
by Kurucz, Bencz$\acute{\mr{u}}$r and Csalog$\acute{\mr
{a}}$ny (\citeyear{Kuruczb07}), it was first addressed in Ruhe
(\citeyear{Ruheb10}) and then
refined by Gabriel and Zamir (\citeyear{Gabrielb5}). Recently, Liu et
al. (\citeyear{Liub7})
extended the work of Gabriel and Zamir (\citeyear{Gabrielb5}) to use
outlier resistant
regressions instead of simple least squares. Several new EM based
imputation methods have been introduced. In particular, Troyanskaya et
al. (\citeyear{Troyanskayab13}) uses such a method to impute missing
values into microarray
experiments while using the SVD to obtain relevant eigen-genes and
eigen-arrays. For further discussion on EM type estimators and a more
complete review of the literature, see Kurucz, Bencz$\acute{\mr{u}}$r
and Csalog$\acute{\mr
{a}}$ny~(\citeyear{Kuruczb07}).

%
%
%

Although an EM type method could certainly be applied in this context,
we choose to use the iterative algorithm of Gabriel and Zamir
(\citeyear{Gabrielb5})
because of its speed in convergence and because we do not wish to make
distributional assumptions about the data. Now, because the signal
variance in our data follows the behavior of the signal itself, we opt
to use a variance weight in the imputation scheme. Of course
initialization of the algorithm is tricky, in particular, when
$w_{ij}=0$, but our use of the first EigenPixel and EigenSignal to
initialize the algorithm proves to work well; see Gabriel and Zamir
(\citeyear{Gabrielb5}) for more discussion on initialization.

The premise of our approach is that each cell has a ``true'' $\mr{Ca}^{2+}$
signal and we are not able to observe that signal because
there are only a finite number of pixel values available
to capture it. The details of our implementation are provided below.

Let $\mf{u}$ and $\mf{v}$ be the first EigenPixel and EigenSignal
associated with the second SVD used to extract the putative $\mr
{Ca}^{2+}$ signal,
which includes saturated pixels, so that $\mf{u}$ and $\mf{v}$ comprise
most of the pixel and signal information of some cell. Continuing with
the example from the previous section, the matrix of interest is
$X^{\prime}_{2}$. Let the dimensions of the $X^{\prime}_{2}$ be $n
\times m$;
because most of the variation is explained by the first component in
the SVD, the rank one approximation can be obtained by minimizing the error
%
%
\begin{equation}
\sum_{i=1}^{n}\sum_{j=1}^{m}(x'_{2ij}-u_{i}v_{j})^2
\label{eq1}
\end{equation}
with respect to $\mf{u}$ and $\mf{v}$. We also wish to weight each term
in the double summation so that it removes the influence of saturated
pixels and takes into account the appropriate variation. We let the
weights be $w_{ij}=\mr{I}_{ij}/(u_{i}v_{j})^2$, where $\mr{I}_{ij}=0$
when $x'_{2ij}=255$, that is, pixel is saturated,\vspace*{1pt} and $\mr{I}_{ij}=1$,
otherwise. Beckers and Rixen~(\citeyear{Beckersb3}) proposed using an indicator to deal
with missing data. We supplement this approach by using
$(u_{i}v_{j})^2$ to scale the term in the summation of (\ref{eq1}) so
that the variance is no larger than 1. Now our new minimization problem becomes
%
%
\begin{equation}
\sum_{i=1}^{n}\sum_{j=1}^{m}w_{ij}(x'_{2ij}-u_{i}v_{j})^2.
\label{eq2}
\end{equation}

We solve the minimization by alternating between $u_{i}$ and $v_{j}$.
Fixing $j$, we can expand the expression in (\ref{eq2}), let
$A_{j}(\mf
{u})=\sum_{i}\mr{I}_{ij}(x'_{2ij}/u_{i})^2$ and $B_{j}(\mf{u})=\sum
_{i}\mr{I}_{ij}(x'_{2ij}/u_{i})$ and we get that $v'_{j}=A_{j}(\mf
{u})/B_{j}(\mf{u})$ solves that portion of the minimization. Similarly,
if we fix $i$, $u'_{i}=A_{i}(\mf{v})/B_{i}(\mf{v})$, where $A_{i}(\mf
{v})=\sum_{j}\mr{I}_{ij}(x'_{2ij}/v_{j})^2$ and $B_{i}(\mf{v})=\sum
_{j}\mr{I}_{ij}(x'_{2ij}/v_{j})$. The new proposed EigenPixel and
EigenSignal vectors are
$\mf{u}^{\mathrm{new}}=\mf{u}'/\|\mf{u}'\|$ and $\mf{v}^{\mathrm
{new}}=\mf{v}'/\|\mf{v}'\|
$ respectively. This gives us a recurrence relation that we can use to
obtain a clearer version of the EigenPixel and EigenSignal, where the
EigenSignal will represent the clarified $\mr{Ca}^{2+}$ signal of interest.
Beckers and Rixen (\citeyear{Beckersb3}) offer a similar recurrence as
a way of
imputing missing values in oceanographic data. We change the number of
relevant components in the SVD and add a weight that includes a
rescaling factor $1/(u_{i}v_{j})^2$. We include this variance rescaling
factor because the variance of the signal and the signal are
synchronized and we want to account for that effect. The pseudo code
used to program this is shown below:

\begin{enumerate}[]
\item Let $\mf{u}^{0}$ and $\mf{v}^{0}$ be the initial EigenPixel and
EigenSignal vectors obtained by taking the SVD of the pixel-wise matrix
that comprises all the pixel and signal information about the cell of
interest, including saturated values.
\item The first proposed EigenPixel and EigenSignal are $\mf
{u}^{1}=\mf
{u}'/\|\mf{u}'\|$ and $\mf{v}^{1}=\mf{v}'/\|\mf{v}'\|$ respectively,
where $u'_{i}=A_{i}(\mf{v}^{0})/B_{i}(\mf{v}^{0})$ and
$v'_{j}=A_{j}(\mf
{u}^{0})/B_{j}(\mf{u}^{0})$.
\item The $(k+1){\mathrm{st}}$ proposed EigenPixel and EigenSignal are
$\mf
{u}^{k+1}=\mf{u}'/\|\mf{u}'\|$ and $\mf{v}^{k+1}=\mf{v}'/\|\mf
{v}'\|$
respectively,\vspace*{-2pt} where $u'_{i}=A_{i}(\mf{v}^{k})/B_{i}(\mf{v}^{k})$ and
$v'_{j}=A_{j}(\mf{u}^{k})/B_{j}(\mf{u}^{k})$.
\item Iterate until convergence.
\end{enumerate}

The missing values are imputed by the corresponding $u_iv_j$ after the
convergence of the algorithm. We check to make sure that any imputed
value for initially saturated pixels do not fall below its saturated
value. In our application of the algorithm to the real data, all
imputed values passed this test. Very rare numbers of pixels experience
total saturation, that is, $\mr{I}_{ij}=0$ for all $j=1,\ldots,512$ given
some fixed $i$. This is particularly true if we only consider a
subinterval of the 85 minute run. Since it is not possible to impute
values for such pixels, they are dropped from our analysis to avoid
creating bias.

Consider $X^{\prime}_{2}$, the $n \times m$ pixel-wise matrix of $n$ pixels
and $m$ time points. For a~fixed pixel $i$, $x'_{2ij}$ has a total of
$m$ potential saturated values. If $\mr{I}_{ij}$ is the indicator
described above and if we let ${p}_{i}=\sum^{m}_{j=1}\mr{I}_{ij}$ be
the number of nonsaturated values for the $i$th pixel across time, then
we made it a rule to remove the $i$th pixel if $\lfloor
{p}_{i}/m\rfloor< 1/8$. This means that we remove any pixel row of the
matrix $X$ if more than 7/8 of it is saturated.

%
\begin{figure}

\includegraphics{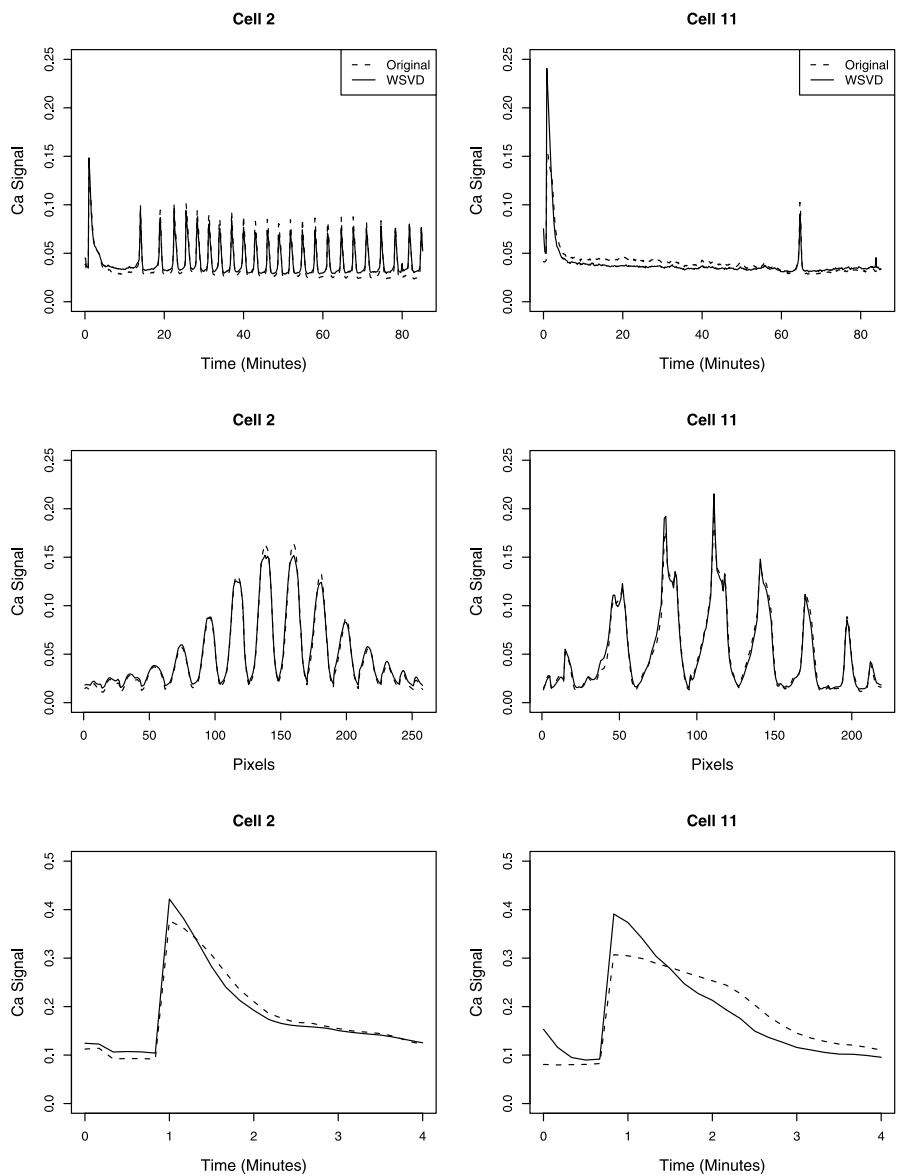}

\caption{Original and WSVD 1st EigenPixel and EigenSignal Vectors for
cell 2
(left column) and cell 11 (right column).
}\label{fig:wsvd2and11}
\end{figure}

\subsection{Application of the WSVD to simulated data}

To evaluate the accuracy of our method, we applied it to a simulated
data set where we used sine curves to emulate the behavior of typical
cell data as shown in Figure \ref{fig:rough2}. Our simulated data
matrix, and first EigenSignal and EigenPixel are shown in Figure \ref
{fig:wsvd2and11}. The data shown in Figure \ref{fig:wsvd2and11} represent the true
signal we are trying to recover. Real data, however, have noise and
also possess saturated pixels that dampen the signal. To duplicate this
behavior, we threshold the data matrix so that everything larger than
$0.50$ is replaced by $0.50$; this mimics a saturation at pixel
locations that have values larger than $0.50$. To add noise, we
introduce realizations from a Gaussian distribution with mean 0 and
variance proportional $(u_{i}v_{j})^2$. We introduced this variance
into the simulated data because it is consistent with the type of
variance observed in the real data and we wanted to emulate that
behavior. Figure \ref{fig:sdp} shows the original and saturated first
EigenPixel and EigenSignal curves. The first EigenSignal and EigenPixel
vectors from the saturated data are both dampened and exaggerated in
different regions.


After applying the weighted SVD to the saturated data matrix, we see
that upon convergence of the algorithm the resulting first EigenPixel
and EigenSignal both come very close to the original curves. The
relative error at every pixel and time point are shown on the right
panel of Figure \ref{fig:sdp}. When we compare the ratio of the error
sum of the saturated over the WSVD Eigen vectors, we see an 11 fold
difference in the EigenPixel and a 8 fold difference in the
EigenSignal. To see the effect of our weight on the results, we removed
the weights and repeated the analysis. Comparing the ratio of the error
sum of the saturated over the WSVD Eigen vectors with no weights, only
a 1 fold difference in the EigenPixel and a~1~fold difference in the
EigenSignal are observed.
%

\subsection{Application of the WSVD to actual data}
For illustration we apply the weighted SVD to the pixel-wise matrices
of cell 2 from the treated group of low hormone level and cell 11 from
the control group of high hormone level. Figure~\ref{fig:wsvd2and11}
shows the first EigenPixel and EigenSignal of both cell 2 and cell 11.
We see that in the peak region of both, between 0--4 minutes into the
experiment, there is a~large difference in the EigenSignal vectors,
especially for cell 11. This is not surprising since most of the
saturation occurs in the peak region of the experiment, hence, pixel
imputation will mostly affect this region. To further explore this
phenomenon, we apply the WSVD only in the peak region (0--4 minutes) and
results are also shown in Figure \ref{fig:wsvd}. We see that the
saturated pixels were dampening the expression in the peak region. This
is a key finding since it is believed that $\mr{Ca}^{2+}$ expression
in this peak
region could be used to characterize cells studied.

We have shown that the weighted SVD can be used to clarify
the $\mr{Ca}^{2+}$ signal in the cells presented. This is an important step
when harnessing $\mr{Ca}^{2+}$ expression from these cells, especially
because $\mr{Ca}^{2+}$ expression is dampened drastically if we do not take
into account the saturation effect.

%
\section{Comparison of $\mr{Ca}^{2+}$ signals: Control and treated}
\label{sec:comp}

\subsection{Initial analysis}

Experience of the third and fourth authors led us to
believe that the $\mr{Ca}^{2+}$ expression observed immediately after
oxytocin exposure is indicative of cell behavior and can
predict the response to a given treatment. This leads us
to consider use of the ``peak'' $\mr{Ca}^{2+}$ signal and the ``post
peak'' $\mr{Ca}^{2+}$ signal, where the ``peak'' signal is obtained
by recovering the signal from the region in the first 4~minutes of the experiment, and the ``post peak'' $\mr{Ca}^{2+}$ signal
is harnessed from the region 40--80 minutes after the
experiment had begun. One goal is to compare how predictive
the initial ``peak'' $\mr{Ca}^{2+}$ signal is compared to the ``post
peak'' $\mr{Ca}^{2+}$ signal. In addition, we have the crucial questions
(a) how does TCDD affect the cells over all, and (b) how
is this response affected by each of the hormone levels in question?

\begin{figure}

\includegraphics{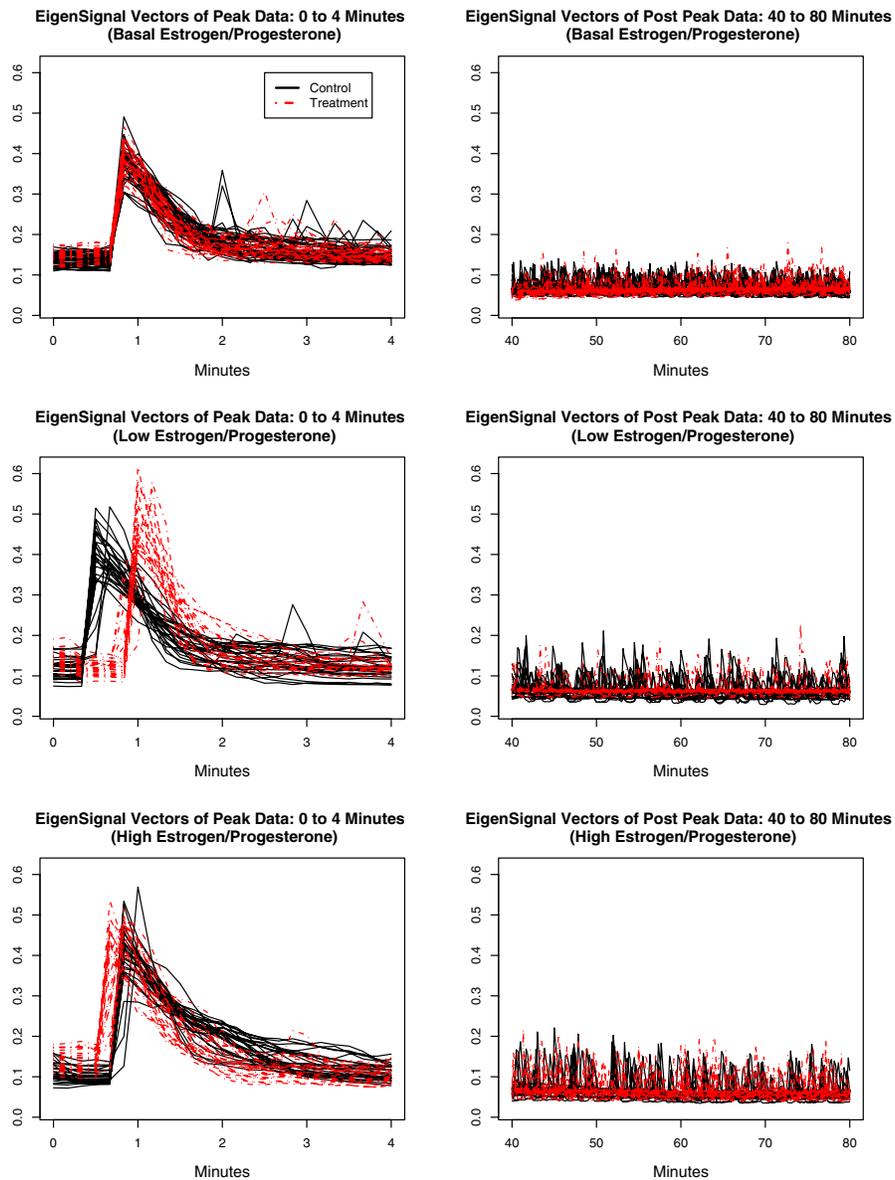}

\caption{\baselineskip=10pt First EigenSignal vectors of the ``peak''
signal (left column ) and the ``post peak'' signal (right column)
obtained from control and treated cells for the three levels of
hormone: basal, low and high. }
\label{fig:fes}
\end{figure}

We first take the weighted SVD as described in the previous section and
plot the first EigenSignal for every cell and for the ``peak'' and
``post peak'' regions; see Figure \ref{fig:fes}. The first thing to
note is that it is easiest to distinguish between the control and
treated cell in the low group. The peaks of the first EigenSignals in
the low hormone cells do not coincide, so it is quite easy to tell the
two groups apart there. About half of the peaks in the high hormone
group coincide and all of the peaks of the control and treatment first
EigenSignals in the basal hormone group coincide. It is much more
difficult to see differences between the control and treated cell lines
in the ``post peak'' region.

%

\subsection{EigenCells}
We next show how to use the SVD a fourth time to extract the effect of
the treatment given to the myometrial cells.

\begin{figure}

\includegraphics{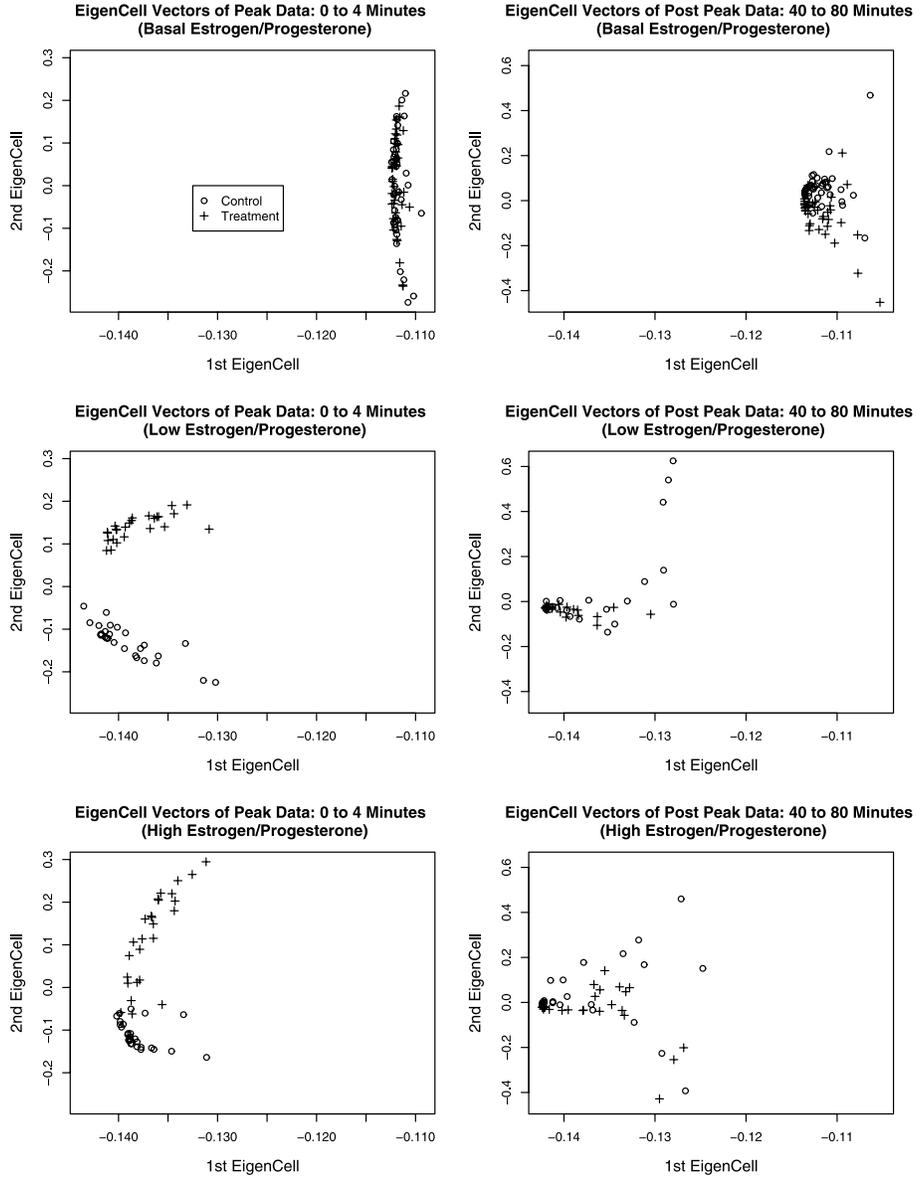}

\caption{\baselineskip=10pt
A scatter plot of the first and second EigenCell vectors for the ``peak''
region (left column) and the``post peak'' region (right column).
The control `$\circ$' and treated `$+$' groups are shown for the three
levels of hormone: basal, low and high.
}
\label{fig:2ndsvd}
\end{figure}

One of our goals is to identify differences, if they exist,
between control and treated cells.
There are many ways in which this comparison can be performed,
but we introduce a new way of distinguishing between these
two levels of drug. Our approach simply performs an additional
SVD on the set of first EigenSignal vectors obtained from the
WSVD. Each cell in the experiment is represented by a first
EigenSignal vector as shown in Figure \ref{fig:fes} and we
combine the first EigenSignal vectors of both the treated cells
and nontreated cells into three matrices, one per hormone level: basal,
low and high. Finally we perform a standard SVD and obtain single value
summary points for each cell, or EigenCell values. Each of the three
hormone levels correspond to a collection of these one point summaries
for a group of cells, which we will call the EigenCell vector. Figure
\ref{fig:2ndsvd} shows the resulting scatter plots of the first two
EigenCell vectors. Because almost 100\% of the variance is explained by
the first two components, we choose to plot only these two. Notice how
easy it is to distinguish between the control and treatment groups in
the ``peak region'' for the low and high hormone level. It is much more
difficult to separate the control and treatment groups in the ``peak''
basal hormone level and in all the hormone levels of the ``post peak'' region.

This clearly shows that the onset of the peak $\mr{Ca}^{2+}$ signal in control
cells is highly organized and occurs immediately following the oxytocin
stimulus in control cells; see Figure \ref{fig:fes}. In the case
of TCDD treated cells that where cultured in low hormone, there is a
delayed peak in the $\mr{Ca}^{2+}$ signal that is thought to result from
suppression of membrane $\mr{Ca}^{2+}$ channels and
pumps that control the release and/or uptake of intracellular $\mr
{Ca}^{2+}$.
Further, the effects of TCDD on myometrial cells appear to vary as a
function of the level of the oxytocin stimulus.

To verify the validity of these claims, we use a 2-fold
cross-validation scheme where 80\% of the data are used to train the
classifier and 20\% to test it. For each random split of data into
training and test sets, we use a $k$-NN classifier with $k =$ 1--5
nearest neighbors on the training and take the average of the error
rates on the test set. The error rates averaged over 1000 runs of the
cross validation are shown in Table \ref{tab:2fcv}. Notice that the
errors reflect our observations, but also show that the ``post peak''
region could be more useful in the basal hormone level if one tries to
predict between the control and treated cell lines.
%
%
\begin{table}[b]
\caption{
Mean error of 1000 runs of our cross-validation scheme to test the
proper classification of the treated and control cell lines using the
EigenCell vectors
}\label{tab:2fcv}%
\begin{tabular*}{\textwidth}{@{\extracolsep{\fill}}lcc@{}}
\hline
\textbf{Hormone level} & \textbf{Peak region} & \textbf{Post peak
region} \\
\hline
Basal & 53\% & 20\% \\
Low & \phantom{0}0\% & 46\% \\
High & \phantom{0}9\% & 26\% \\
\hline
\end{tabular*}
\end{table}

\subsection{$\mr{Ca}^{2+}$ peak comparison}

\begin{figure}

\includegraphics{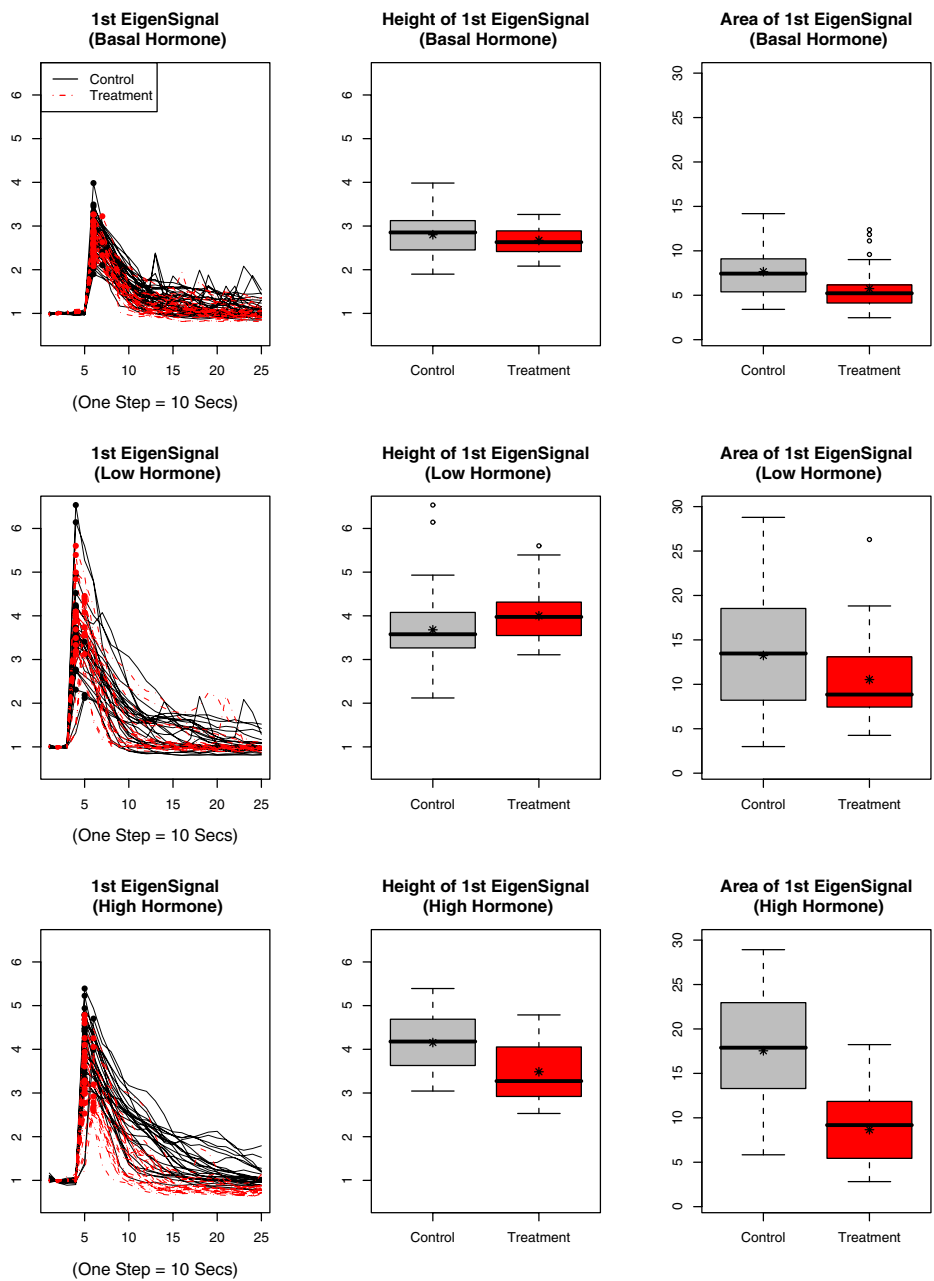}

\caption{\baselineskip=10pt
Left column: Normalized and landmarked first
EigenSignal vectors of the ``peak'' region in the control
and treated cells for each of the three hormone levels:
basal, low and high.
Middle column: Boxplots of the peak height
for the control and treated cells in each of the three hormone
levels: basal, low and high. Right column:
Boxplots of the area in the ``peak'' region for the control
and treated cells in each of the three hormone levels: basal, low and
high.}
\label{fig:bxp}
\end{figure}

Of course, all these results depend on the original structure of the
data, meaning that no manipulation was made to alter the original $\mr
{Ca}^{2+}$
response other than the imputation of values where saturation occurs.
If we wanted to compare the peaks of the initial $\mr{Ca}^{2+}$ signal
directly,
we would have to align the peaks by normalizing them, that is, dividing
the EigenSignal by the first 3 initial values, and also perform
landmark registration, where the landmark would be the point where the
signal begins to rise. Figure \ref{fig:bxp} shows the normalized and
landmarked first EigenSignal curves obtained from the ``peak'' region
after applying the WSVD. A comparison of the peak height and peak area
between control and treatment groups is made. By looking at the
boxplots in Figure \ref{fig:bxp}, it is reasonable to hypothesize that
the mean peak height and area in the control group are greater than
those of the treated group. We performed an exact test where we
permuted the labels of the peak height and peak area 1,000,000 times.
Table \ref{tab:permtest} shows the resulting test statistic and $p$-value
for the peak height and area.

We see a significant difference in the area of the $\mr{Ca}^{2+}$ signals
when comparing the control and TCDD treated cells. This suggests that
TCDD may perturb one or more pathways that regulate $\mr{Ca}^{2+}$
entry through
channels in the plasma
membrane, $\mr{Ca}^{2+}$ release from intracellular stores in the endoplasmic
reticulum (ER) or other mechanisms to remove $\mr{Ca}^{2+}$ from the
cytosol by
pumps in the plasma membrane or ER membrane. Each of these pathways can
now be analyzed in turn to identify the molecular basis for altered
$\mr{Ca}^{2+}$
signals in these cells and, therefore, the physiological relevance of the
decrease in $\mr{Ca}^{2+}$ signaling will be determined. Nevertheless, a
significant alteration in calcium signaling indicates a significant
change in the myometrial cell contractile response.

The differences in peak height, however, seem to be a bit mixed. One
important result to note is that for both the peak height and peak area
the permutation test is highly significant in the high hormone group,
indicating a strong difference between the control and treated cell lines.
A decrease in $\mr{Ca}^{2+}$ signaling corresponds with a decrease in
myometrial
contraction (i.e., uterine contraction), and a high level of
estrogen/progesterone hormone level in myometrial cells is meant to
simulate a response of these cells at the late stages of pregnancy.
This means that ``normal'' function of the uterus could be compromised
by TCDD at the late stages of pregnancy, an important finding that
deserves further investigation and expansion of our research.
%
%
\begin{table}
\caption{
Test statistic (difference of mean) between control and treated peak
height and peak area and p-values after running permutation tests with
1,000,000 permuted samples
}
\label{tab:permtest}
\begin{tabular*}{\textwidth}{@{\extracolsep{\fill}}lccc@{}}
\hline
\textbf{Hormone level} & \textbf{Test stat./$\bolds{p}$-value} &\textbf{Peak
height} & \textbf{Peak
area} \\
\hline
{Basal} & $\bar{x}_{C}-\bar{x}_{T}$ & 0.129 & 1.906\\
& $p$-value & 0.074 & 0.001\\
{Low} & $\bar{x}_{C}-\bar{x}_{T}$ &$-0.32$\phantom{00.} &
2.71\phantom{0}\\
& $p$-value &0.902 & 0.046\\
{High} & $\bar{x}_{C}-\bar{x}_{T}$ &0.669 & 8.883\\
& $p$-value &0.000 & 0.000\\
\hline
\end{tabular*}
\end{table}

%

%


\section{Conclusion}
\label{sec:conclusion}

In this work we use the SVD in four different ways:
\begin{enumerate}[]
\item First, we use it to detect the $\mr{Ca}^{2+}$ signal by using the
initial first EigenPixel vector. This approach summarizes
cell location information across all 512 images instead of
using only one image as is typically done for these data.
\item Second, another SVD was then used to extract the $\mr{Ca}^{2+}$
signal from the pixel-wise matrix derived after segmenting the cell
region in raw images. These first EigenSignal and first EigenPixel
vectors serve as the templates used to ``clean up'' the signal.
\item Third, we used those candidate EigenSignal and EigenPixel vectors
to clarify the $\mr{Ca}^{2+}$ signal by applying a new weighted SVD,
the WSVD, to
impute values where saturation occurs in the signal.
\item Finally, we use the singular value decomposition once more to
discriminate between control and treated EigenSignal vectors resulting
from the WSVD. We summarize the variation in the control and treated
cell lines by capturing the variability of each cell into one value per
cell, giving us the EigenCell vector.
\end{enumerate}
To our knowledge $\mr{Ca}^{2+}$ signal detection, extraction,
clarification and comparison using the SVD has not been previously
performed. These four applications of the singular value decomposition
to analyze $\mr{Ca}^{2+}$ signaling in myometrial cells show its
utility and
flexibility for analyzing complex $\mr{Ca}^{2+}$ signals such as oscillations
and waves.

An additional finding is that saturation undermines the $\mr{Ca}^{2+}$ signal
obtained by simply averaging the pixels representing the cell.
Correcting the effects of saturation must be an integral step
while studying these type of data. Moreover, the hypothesized
importance of the ``peak'' region as being a way of characterizing
cells of this type seems to be a valid claim. From our analysis
we were able to clearly distinguish between the treated and control
groups by using the area in the ``peak'' region and by using the
scatter plots of EigenCells vectors obtained in our fourth and
final application of the SVD.

To conclude, we have shown the importance of the initial peak
in $\mr{Ca}^{2+}$ signaling of myometrial cells by the SVD, and also exhibit
new uses of the SVD to segment, extract, clarify and compare $\mr{Ca}^{2+}$
signals in this context.


%

\begin{supplement}[id=suppA]
\sname{Supplement A}
\stitle{Calcium ion signaling movies with TCDD exposure}
\slink[doi]{10.1214/07-AOAS253SUPPA}
\sdatatype{.zip}
\slink[url]{http://lib.stat.cmu.edu/aoas/253/dir2\_T.zip}
\sdescription{When unzipped, the movie is
in .avi format, and is 30 MB in size. One can view it, for
example, using windows media player.}
\end{supplement}

\begin{supplement}[id=suppB]
\sname{Supplement B}
\stitle{Calcium ion signaling movies without TCDD exposure}
\slink[doi]{10.1214/07-AOAS253SUPPB}
\sdatatype{.zip}
\slink[url]{http://lib.stat.cmu.edu/aoas/253/dir2\_C.zip}
\sdescription{When unzipped, the movie is
in .avi format, and is 40 MB in size. One can view it, for
example, using windows media player.}
\end{supplement}
%

\printaddresses

\end{document}